\documentclass[12pt,a4paper]{article}
\usepackage{amsmath}
\usepackage{amsfonts}
\usepackage{amssymb}
\begin{document}
	
\author{Doru Constantin\footnote{constantin@unistra.fr}}
\date{Institut Charles Sadron, CNRS and Universit\'e de Strasbourg, 67034 Strasbourg, France}
	\title{Simple derivation of de Gennes narrowing}
	\maketitle
	\begin{abstract}
	I present a simple derivation of the de Gennes narrowing phenomenon.
	\end{abstract}
	
In colloidal solutions, a widely-used relation connects the scale-dependent collective diffusion constant and the structure factor:
\begin{equation}
D_c(q) =\frac{D_0}{S(q)}
\label{eq:dGn}
\end{equation} and is generally known as de Gennes narrowing since its use by de Gennes in the context of quasi-elastic neutron scattering from liquids \cite{deGennes}.

Intuitively, equation \eqref{eq:dGn} makes sense: if \(S(q)\) is high for a certain value of \(q\) then fluctuations with that particular \(q\) are frequent, meaning that their energetic cost is low and that they will decay slowly (in the time domain), meaning that the associated mode (in the frequency domain) is narrow, hence the name. However, I have not yet found in the literature a simple derivation: this is what I will attempt below.

Consider a system characterized by a conserved scalar parameter \( \phi(\mathbf{r})\) (for instance, the local particle concentration in a suspension \( \phi(\mathbf{r}) = \rho(\mathbf{r}) - \rho_0\), with \(\rho_0\) the equilibrium value). We are interested in the excess free energy due to inhomogeneities of this field: \(\mathcal{F} - \mathcal{F}_0 = \mathcal{F}[\phi(\mathbf{r})]\) or, in terms of its Fourier components: \(\mathcal{F} - \mathcal{F}_0 = \mathcal{F}[\phi(\mathbf{q})]\).

For an isotropic system in the absence of applied fields, all fluctuations \( \phi(\mathbf{q})\) with \( \left |\mathbf{q} \right | > 0\) will eventually decay to zero. To fix the ideas, we will consider an overdamped relaxation (model B in the Hohenberg-Halperin classification \cite{Hohenberg}, but of course far from criticality).

Let us write Fick's laws, with \( \mathbf{j} \) and \( \mu\) the current and chemical potential associated to \( \phi\) (this is similar to the presentation in \cite{Hohenberg}, Eqs.~(2.2)-(2.9)):
\begin{equation}
\label{eq:defs}
\frac{\partial \phi(\mathbf{r},t)}{\partial t} = - \nabla \mathbf{j} \, ; \quad \mathbf{j} = - \lambda \nabla \mu \, ; \quad \mu = \frac{\delta \mathcal{F}}{\delta \phi(\mathbf{r},t)} \, \Rightarrow \, \frac{\partial \phi(\mathbf{r},t)}{\partial t} = \lambda \nabla ^2 \frac{\delta \mathcal{F}}{\delta \phi(\mathbf{r},t)}\end{equation} where \( \delta\) denotes the functional derivative. The second relation in \eqref{eq:defs} serves as a definition for the transport coefficient \(\lambda\).
Introducing the Fourier components yields: \begin{equation}
\label{eq:relax}
\frac{\partial \phi(\mathbf{r},t)}{\partial t}= \lambda \nabla ^2 \frac{\delta \mathcal{F}}{\delta \phi(\mathbf{r},t)}  \Rightarrow \frac{\partial \phi(\mathbf{q},t)}{\partial t} = - \lambda q^2 \frac{\text{d} \mathcal{F}}{\text{d}\phi(\mathbf{q},t)}
\end{equation} If the different Fourier modes are uncoupled, we can write the equipartition relation:
\begin{equation}
\label{eq:equi}
\mathcal{F} = \mathcal{F}_0 \sum_{\mathbf{q}}  \frac{A(\mathbf{q})}{2} |\phi(\mathbf{q})| ^2 \Rightarrow \frac{\text{d} \mathcal{F}}{\text{d}\phi(\mathbf{q},t) } = A(\mathbf{q}) \phi(\mathbf{q})
\end{equation} Let us introduce the time-dependent structure factor \begin{equation}
\label{eq:Sqt}
S(\mathbf{q},t) = \left \langle \phi(\mathbf{q},t) \phi(\mathbf{-q},0) \right \rangle
\end{equation} with \( \left \langle \cdot \right \rangle\) the ensemble average. 

Plugging \eqref{eq:equi} into \eqref{eq:relax}, multiplying by \(\phi(-\mathbf{q},0)\) and averaging yields the evolution of mode \(\mathbf{q}\): \begin{equation}
\label{eq:evol}
\frac{\partial}{\partial t} \left \langle \phi(\mathbf{q},t) \phi(-\mathbf{q},0) \right \rangle =- \lambda q^2 A(\mathbf{q}) \left \langle \phi(\mathbf{q},t) \phi(-\mathbf{q},0) \right \rangle \Rightarrow \frac{S(\mathbf{q},t)}{S(\mathbf{q},0)} = \exp \left \lbrace - D_c(q) q^2 t \right \rbrace
 \end{equation} where we invoked the isotropy of the system. The collective diffusion coefficient is given by: \begin{equation}
\label{eq:Dc}
D_c(q) = \lambda A(q)
 \end{equation} On the other hand,equipartition also implies: \begin{equation}
S(\mathbf{q}) = S(\mathbf{q},0) = \left \langle \phi(\mathbf{q},0) \phi(\mathbf{-q},0) \right \rangle = \frac{k_B T}{A(\mathbf{q})} .
\label{eq:Sq}
\end{equation} From \eqref{eq:Dc} and \eqref{eq:Sq} we finally obtain:
\begin{equation}
\label{eq:final}
D_c(q) = \frac{\lambda k_B T}{S(q)}
\end{equation}

\end{document}